\documentclass[%
 reprint,
 amsmath,
 amssymb,
 aps,
]{revtex4-2}

\usepackage{graphicx}
\usepackage{dcolumn}
\usepackage{bm}
\usepackage{color}

\begin{document}

\preprint{APS/123-QED}

\title{
High-power laser experiment on developing supercritical shock propagating
in homogeneously magnetized \textcolor{black}{plasma of ambient gas origin}}

\author{S. Matsukiyo$^1$}
 \email{matsukiy@esst.kyushu-u.ac.jp}
 \altaffiliation[Also at ]{ICSWSE, Kyushu University.}
\author{R. Yamazaki$^2$}%
\author{T. Morita$^1$}%
\author{K. Tomita$^{3,1}$}%
\author{Y. Kuramitsu$^4$}%
\author{T. Sano$^5$}%
\author{S. J. Tanaka$^2$}%
\author{T. Takezaki$^{6,7}$}%
\author{S. Isayama$^1$}%
\author{T. Higuchi$^8$}%
\author{H. Murakami$^8$}%
\author{Y. Horie$^8$}%
\author{N. Katsuki$^8$}%
\author{R. Hatsuyama$^8$}%
\author{M. Edamoto$^8$}%
\author{H. Nishioka$^8$}%
\author{M. Takagi$^8$}%
\author{T. Kojima$^8$}%
\author{S. Tomita$^{9,10}$}%
\author{N. Ishizaka$^2$}%
\author{S. Kakuchi$^2$}%
\author{S. Sei$^2$}%
\author{K. Sugiyama$^2$}%
\author{K. Aihara$^2$}%
\author{S. Kambayashi$^2$}%
\author{M. Ota$^{5,11}$}%
\author{S. Egashira$^{11}$}%
\author{T. Izumi$^{11}$}%
\author{T. Minami$^{12}$}%
\author{Y. Nakagawa$^{12}$}%
\author{K. Sakai$^{12}$}%
\author{M. Iwamoto$^{1,13}$}%
\author{N. Ozaki$^4$}%
\author{Y. Sakawa$^5$}%
\affiliation{%
 $^1$Faculty of Engineering Sciences, Kyushu University
}%
\affiliation{%
 $^2$Department of Physics and Mathematics, Aoyama Gakuin University}%
\affiliation{%
 $^3$Division of Quantum Science and Engineering, Hokkaido University}
\affiliation{%
 $^4$Graduate School of Engineering, Osaka University}%
\affiliation{%
 $^5$Institute of Laser Engineering, Osaka University}
\affiliation{%
 $^6$Faculty of Engineering, University of Toyama}
\affiliation{%
 $^7$Department of Creative Engineering, National Institute of Technology, Kitakyushu College}
\affiliation{%
 $^8$Interdisciplinary Graduate School of Engineering Sciences, Kyushu University}
\affiliation{%
 $^{9}$Astronomical Institute, Tohoku University}
\affiliation{%
 $^{10}$Frontier Research Institute for Interdisciplinary
Sciences, Tohoku University}
\affiliation{%
 $^{11}$Graduate School of Science, Osaka University}
\affiliation{%
 $^{12}$Graduate School of Engineering, Osaka University}
\affiliation{%
 $^{13}$Department of Earth and Planetary Science, University of Tokyo}

\date{\today}

\begin{abstract}
\textcolor{black}{A \textcolor{black}{developing} supercritical
collisionless shock propagating in a homogeneously magnetized
\textcolor{black}{plasma of ambient gas origin having higher uniformity than the
previous experiments} is formed by using high-power laser
experiment. The \textcolor{black}{ambient} plasma is not contaminated by
the plasma produced in the early time after the laser shot.
While the observed \textcolor{black}{developing shock does not
have stationary downstream structure}, it possesses
some characteristics of a magnetized supercritical shock,
which are supported by a one-dimensional full
particle-in-cell simulation taking the effect of finite
time of laser-target interaction into account.}
\end{abstract}

\maketitle


\textcolor{black}{\section{\label{sec:intro}Introduction}}

\textcolor{black}{
Magnetized collisionless shocks are ubiquitous and
believed to play important roles in a variety of space
and astrophysical phenomena such as particle
acceleration and heating, coherent wave emission,
amplification of magnetic field, etc. Explosive
phenomena in space such as a supernova remnant and a
solar flare often accompany one or more shocks (e.g.,
\cite{ghavamian13}). A stellar wind usually accompany a
termination shock and planetary bow shocks
\cite{burgess15}. There also exist large scale shocks
like galaxy cluster merger shocks \cite{frassati22}.
}

\textcolor{black}{
Recent advance of laboratory experiment on collisionless
shocks using high-power laser enables us to study the
spatio-temporal evolution of the system experimentally.
A magnetized subcritical shock experiment using a high-power
laser was first conducted by
\cite{niemann14,schaeffer14,schaeffer15}. The first high-power
laser experiment of a magnetized supercritical
shock formed in counter streaming plasma was performed
by \cite{kuramitsu16}, although they focused on the
Richtmyer-Meshkov instability rather than the shock
itself. Here, a magnetized shock with the Alfv\'{e}n
Mach number greater than a critical value (roughly $\sim
3$) is called a supercritical shock in which reflected
ions play a significant role in energy dissipation.
Recently, \cite{yao22,yao21,schaeffer19,schaeffer17} proposed a
generation method of a magnetized supercritical shock
and observed the plasma interaction evolving into a
shock.
}

\textcolor{black}{
One of the difficulties of the shock experiment is to
produce a shock in a homogeneously magnetized uniform medium as
in space. In a number of past experiments
two solid targets are irradiated so that the interaction
between two ablation plasmas results in shock formation
\cite{kuramitsu16,schaeffer17,schaeffer19}. In this case the two
plasmas may be partially magnetized due to laser induced
magnetic field (Biermann battery effect). But the
generated magnetic field is highly varying in both space
and time. Also, the two plasmas may not be uniform.
\cite{yao21,yao22} used a gas nozzle to form an ambient
plasma which may possess higher uniformity. We utilize
here a method to generate a
supercritical shock in a \textcolor{black}{homogeneously magnetized
ambient plasma at rest}. The method was developed by
\cite{yamazaki20}. In this platform an ambient magnetic
field is applied by using a Helmholtz coil for
sufficiently long time and in sufficiently large volume
filled with a background gas through an experiment. By
irradiating a solid target with a laser, a target is
ablated and the surrounding gas filling
an entire chamber is ionized by the strong
radiation emitted by the laser-target interaction.
\textcolor{black}{We call this ionized gas as a gas plasma
in this paper. Although the gas is uniform, the uniformity
of the gas plasma is not perfect due to varying/uncontrolled
ionization level. Even though, as discussed in
\cite{yamazaki20}, the gas plasma in the region
of interest produced in this way may be more uniform than the
ambient plasma in the past experiments. In particular,
if the gas plasma is fully ionized ($Z>1$, where $Z$
denotes an averaged valence of ions), the Alfv\'{e}n
velocity of unshocked gas plasma is uniform.} A
target plasma plays a role of piston to form a shock in
\textcolor{black}{an ambient gas plasma}. Here, we focus on
longer time evolution of the magnetized \textcolor{black}{developing} shock than
\cite{yamazaki20}, $t > 30$ ns, where the observed
plasma is not contaminated by the target plasma.
}

A spatio-temporal evolution of the system was observed
by optical measurements such as self-emimssion streaked
optical pyrometer (SOP) and collective Thomson
scattering (CTS). The parameters of an unshocked
upstream gas plasma were determined. The time evolution
of the shock is compared with a one-dimensional full
particle-in-cell (PIC) simulation.

\textcolor{black}{\section{\label{sec:setting}Experimental Settings and measurement system}}

%
\begin{figure}[ht]
\includegraphics[clip, width=0.95\columnwidth]{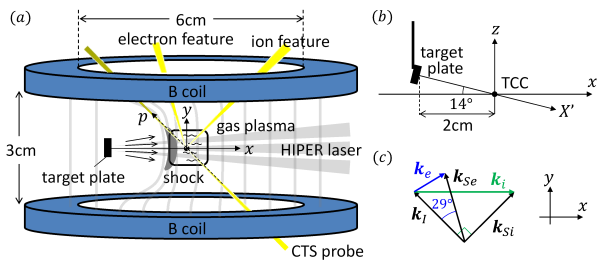}
\caption{\label{fig:expsetting}Experimental settings:
(a) top view, (b) side view near the target, and (c) the
relation among the wavenumbers of incident probe light
($\mbox{\boldmath $k$}_I$), scattered waves \textcolor{black}{($\mbox{\boldmath $k$}_S$)},
and plasma waves \textcolor{black}{($\mbox{\boldmath $k$}$)} for CTS measurement.}
\end{figure}
%

The experiment was carried out with Gekko XII HIPER
laser facility at the Institute of Laser Engineering,
Osaka University. Four beams of long pulse laser [energy
of $\sim 700$ J at 1053 nm, Gaussian pulse with 1.3 ns
duration and 2.8 mm spot, F/15 for each beam] irradiate
an aluminum (Al) plate target with 2 mm thickness at
$x=-2$ cm, behind the target chamber center (TCC), to
create a high speed plasma flow normal to the target
plate (Fig.\ref{fig:expsetting}). 5 Torr nitrogen (N)
gas filling the target environment was ionized due to
the radiations from the laser-target interaction.
\textcolor{black}{The
target position is more distant from the TCC than
\cite{yamazaki20} ($x = -1.4 \cos{14^{\circ}}$ cm) so as
to capture longer time evolution of the system, although
the structures associated with earlier time phenomena
are outside the field of view.}
The ambient magnetic
field parallel to the target surface was applied by
using a Helmholtz coil (6 cm diameter of each coil and 3
cm inter distance) driven by a portable pulsed magnetic
field generation system \citep{edamoto18} so that the
gas plasma was \textcolor{black}{homogeneously} magnetized. Two sets of
capacitor banks, each consists of four condensers with 3
mF capacitance, were charged with $1.4$ kV and $\sim 3$
kA current was applied with a pulse duration of over 100
$\mu$s. This enabled us to apply $\sim 3.8$ T ambient
magnetic field in the region of interest. The high speed
target plasma carrying Biermann battery magnetic field
acts as a magnetic piston to form a shock in the
\textcolor{black}{ambient} gas plasma.

We measured spatio-temporal evolution of the system by
using a 450 nm bandpass filtered SOP. Spatial
information was obtained along the line parallel to the
target normal, defined as $X'$-axis inclined in the
$x-z$ plane by $-14^{\circ}$ from the $x$-axis
[Fig.\ref{fig:expsetting}(b)]. CTS measurement
\textcolor{black}{for ion feature} was
conducted to obtain local plasma quantities at a
particular time \cite{tomita17,morita19}. The probe light
(Nd:YAG laser, wavelength $\lambda_0=532$ nm, pulse
duration 8 ns, laser energy 300 mJ) was injected along
the $p$-axis which is inclined by $45^{\circ}$ from the
$y$-axis in the $x-y$ plane
[Fig.\ref{fig:expsetting}(a)]. The detection system was
placed in the direction of \textcolor{black}{$-90^{\circ}$} from the
$p$-axis, which mainly consists of a handmade
triple-grating spectrometer [spectral resolution was \textcolor{black}{10
pm}] and an intensified CCD camera (Princeton Inst.,
PIMAX4, gate width 5 ns quantum efficiency at
$\lambda_0$ was 40\%). In all the above coordinate
systems, the origin is shared at TCC.

\textcolor{black}{\section{Experimental results}}

\begin{figure}[ht]
\includegraphics[clip, width=1.0\columnwidth]{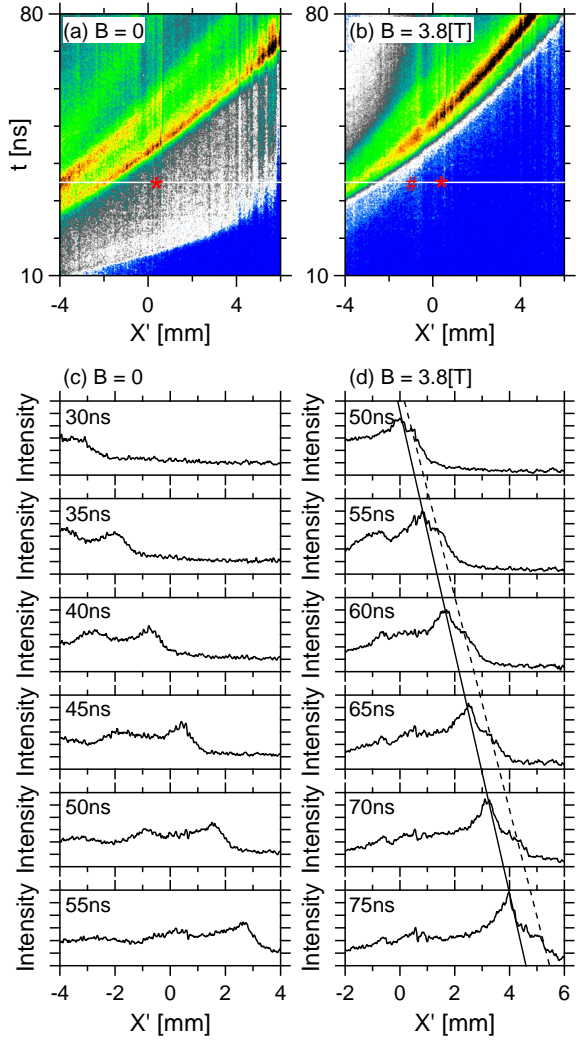}
\caption{\label{fig:sop}SOP data for the cases of (a,c) $B=0$
and (b,d) $B=3.8$ T. In (a) and (b) a vertical axis denotes
an elapsed time from the laser shot. The color shows
self-emission intensity normalized to the far upstream value.
The intensity profiles at different times are plotted in (c)
and (d). The solid and the dashed lines in (d) roughly trace
the positions of the main peak and the plateau-like structure.
CTS spectra at the red markers in (a) and (b) are shown in Figs.\ref{fig:TSspectra}(a) and \ref{fig:TSspectra}(b), at the white marker is shown in Fig. 5(b), respectively.}
\end{figure}
%
The color images of the SOP data show that the peaks of
high intensity emission propagated in the positive
$X^{\prime}$ direction for $B=0$ [Fig.\ref{fig:sop}(a)]
and $B=3.8$ T [Fig.\ref{fig:sop}(b)]. The propagation
speed estimated at $X'=0$ (TCC) was $\sim 270$ km/s in (a)
and $\sim 210$ km/s in (b), respectively. In (a) a region
where the emission intensity was slightly enhanced was
extended in the right (upstream side) of the intensity
peak. \textcolor{black}{This structure was discussed by \cite{yamazaki20}}.
Hereafter, we call this as a precursor. In (b) the similar
precursor with rather smaller spatial extent was seen only
for $t < 40$ ns.
\textcolor{black}{This could be the 'R2' discussed in
\cite{yamazaki20}. They interpreted this as the structure
associated with gyrating N ions which are reflected by
expanding target plasma in the early time phase.
Another possible interpretation is that it is the
structure associated with gyrating target ions. Using the
value of 800 km/s as the initial target Al ion velocity
(according to \cite{yamazaki20}), their gyro radius for
$B=3.8$ T is $\rho_{Al} \sim 58/Z_{Al}$ mm, where $Z_{Al}$
is the average valence of Al ions. Since outermost
electrons of an Al atom are easily excluded, it is natural
to infer that $Z_{Al} \geq 3$ so that $\rho_{Al} < 20$ mm.
Hence, we conclude that the structure observed after $t >
40$ ns is the one formed in a magnetized gas plasma not
contaminated by the gyrating Al and N ions produced in the early time
after the laser shot.} The main intensity peak for
$B=0$ [Fig.\ref{fig:sop}(c)] has relatively simple and
stable structure. In contrast, the intensity peak for
$B=3.8$ T clearly has sub-structures and the details of
the profile varies in time [Fig.\ref{fig:sop}(d)]. In front of
the main peak a local plateau-like structure is formed and
the spatial size of the whole transition region widens in
time during the period shown. The decay length of the
whole transition region reaches $\sim 2$ mm, which is
apparently larger than that ($\lesssim 1$ mm) of the main
peak in the case of $B=0$.

\begin{figure}[ht]
\includegraphics[clip, width=1.0\columnwidth]{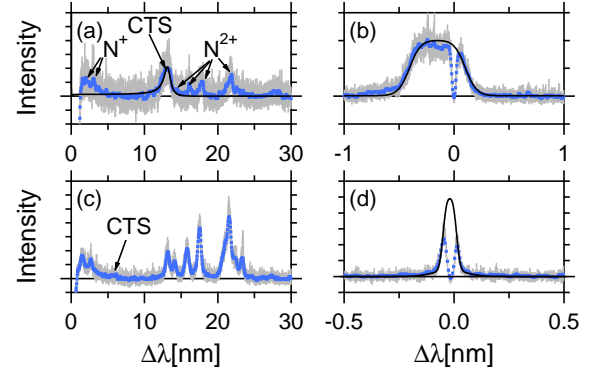}
\caption{\label{fig:TSspectra}\textcolor{black}{CTS spectra
at $t=35 (\pm 2.5)$ ns for (a,c) $B=0$ and (b,d) $B=3.8$ T.
In the panels (a) and (b), the vertical and the horizontal axes
indicate the position along the probe light ($p$-axis) and the
wavelength shift, $\Delta \lambda$, from $\lambda_0$, the color
denotes the arbitrary intensity of the scattered signal,
respectively. The corresponding time integrated to obtain this
data is shown as the red square in Fig.\ref{fig:sop}(a) and (b).
Assuming that a planar structure is propagating
along the $X'$-axis, a position along the $p$-axis is
projected onto the $X'$-axis as $X'=-p \cos 14^{\circ} / \sqrt{2}$.
The projection of the position is indicated as the red vertical bars
in Fig.\ref{fig:sop}(a) and (b). In the panels
(c) and (d), the cross-section along the white line ($p=-0.5$ mm)
in (a) and (b) is plotted. See the text in detail.}}
\end{figure}
\textcolor{black}{
The CTS spectrum in the precursor for $B=0$ is shown in
Fig.\ref{fig:TSspectra}(a). The vertical axis denotes
the position, $p$, along the path of the probe light,
while the horizontal axis indicates the deviation of
wavelength from the probe light ($\lambda_0 = 532$ nm).
The spectrum was obtained as a snapshot at $t = 35
(\pm2.5)$ ns. The corresponding time domain (5 ns gate width)
in the SOP data is indicated by the red square in
Fig.\ref{fig:sop}(a). In Fig.\ref{fig:TSspectra}(a),
the signal is cut at around $\lambda_0$ to avoid stray
light. An abrupt shift and broadening of the spectrum
is observed in the region of $p>0$. The peak intensity
occurs in slightly above $p=1$ mm, which may correspond
to the peak in the red squared region in
Fig.\ref{fig:sop}(a). Note that the positive $p$
corresponds to the negative $X'$. The deviation of the
two axes may cause the slight discrepancy of the peak
position. The cross-section of the spectrum in the
precursor at $p=-0.5$ mm is shown in
Fig.\ref{fig:TSspectra}(c). The blue dotted line
denotes an arbitrary intensity averaged over $-0.51
\leq p \leq -0.49$ mm at each wavelength, while the
maximum and the minimum values in the region give the
error bars indicated by the gray vertical lines. To fit
the averaged data, we use the spectral density function
written as
\begin{equation}
    S(\bm{k},\omega) ={2 \pi \over k} \left| 1-{\chi_e \over \epsilon}
    \right|^2 f_{e0}\left( {\omega \over k} \right) + {2 \pi Z_N \over k}
    \left| {\chi_e \over \epsilon} \right|^2 f_{N0} \left( {\omega \over k} \right).
\label{sdfunc}
\end{equation}
Here, $\omega$ and $k$ are the frequency and wavenumber
of the density fluctuations of the plasma, $\epsilon$ and
$\chi_e$ denote electric permittivity and electron
susceptibility, $f_{e0}$ and $f_{N0}$ are the
shifted-Maxwellian distribution functions of electrons
and N ions, respectively. In the experiment the obtained
signal is the convolution of $S(\bm{k},\omega)$ and
$R(x)$, as $\hat{S}(\bm{k},\omega)=\int
S(\bm{k},\omega^{\prime})R(\omega^{\prime}-\omega)d\omega
^{\prime}$, where $R$ denotes the resolution of the
spectrometer evaluated from the Rayleigh scattering. The
electron density $N_e$ is determined by using the
following relation between the intensities of the Thomson
scattering and the Rayleigh scattering, $I_T/I_R =
(N_e/N_{NG})(\sigma_T/\sigma_R)(E_T/E_R)(S_N/2\pi)$.
$N_{NG}$ is the density of nitrogen gas, $\sigma$ the
scattering cross section, $E$ the laser energy. The
subscripts $T$ and $R$ denote Thomson and Rayleigh
scatterings, respectively. $S_N$ is the total intensity
integrated over $\omega$ of the second term of
eq.(\ref{sdfunc}). The black solid line is obtained using
the above method as the best fit using electrons' density
$N_e \approx 1.4\times 10^{18}$ cm$^{-3}$, temperature
$T_e \approx 240$ eV, drift velocity $v_{de} \approx 40$
km/s, ions' temperature $T_N \approx 450$ eV, charge
state $Z_N \approx 3.9$, and drift velocity $v_{dN}
\approx 40$ km/s, respectively. These values are not very
far from those obtained by \cite{yamazaki20} in the
precursor at an earlier time and at a closer position to
the target. Since $N_e/Z_N \approx 3.6 \times 10^{17}$ cm$^{-3}$
is comparable to the density of N atoms contained in a 5 Torr gas
($\approx \textcolor{black}{3.2} \times 10^{17}$ cm$^{-3}$), the local plasma is
considered to be fully ionized.
}

\textcolor{black}{
The CTS spectrum obtained at $t = 35 (\pm2.5)$ ns for
$B=3.8$ T is shown in Fig.\ref{fig:TSspectra}(b). A
gradual broadening and shit of the spectrum occurs in
$p>0.5$ mm. This region may correspond to the edge of the
small precursor in Fig.\ref{fig:sop}(b). The region of
$p<0.5$ mm is regarded as an upstream gas plasma. The
cross-section of the spectrum in this region at $p=-0.5$
mm is shown in Fig.\ref{fig:TSspectra}(d). Despite that
the peak is cut, we fitted the blue dotted line to obtain
the theoretical line shown by the black solid one by
using $N_e \approx 4.5 \times 10^{17}$ cm$^{-3}$, $T_e
\approx 7$ eV, $T_N \approx 5$ eV, $Z_N \approx 1.3$, and
$v_{de} = v_{dN} \approx 5$ km/s, respectively. Note that
the above values of $N_e$ and $Z_N$ indicate that the
5 Torr gas was fully ionized. Using
these values, the Alfv\'{e}n velocity in the magnetized
gas plasma upstream is estimated as $V_{AG} \approx 38.6$
km/s. Hence, the Alfv\'{e}n Mach number corresponding to
the propagation speed of the peak, $210$ km/s, is $M_A
\approx 5.4$ indicating that the shock is supercritical.
The upstream electron beta is $\beta_{e} \approx 0.1$.
}

\textcolor{black}{\section{Numerical simulation and discussions}}

To interpret the above experimental results, we
numerically simulate the interaction between a target
plasma and a gas plasma by using a one-dimensional PIC
simulation. Recently, a method to simulate ambient and
target plasma interaction is proposed by \cite{fox18} and
developed by \cite{schaeffer20}. Here, a similar but
different method is used. Initially the system is filled
with a uniform thermal gas plasma consisting of
electrons and ions. The valence of gas ions is assumed to
be $Z_G=1$, based on the CTS analysis ($Z_N=1.3$). The
ion-to-electron mass ratio is $m_G/m_e=100$. The ratio of
electron plasma frequency to cyclotron frequency is
$\omega_{Ge}/\Omega_{Ge}=20$. The electron and ion betas
are $\beta_{Ge}=\beta_{Gi}=0.1$ \textcolor{black}{and
their distribution function is Maxwellian}. At
$X'=-2$ cm$/\cos
14^{\circ}$, a dense target plasma is injected during $0
\le \Omega_{Gi}t \le 0.035$, where $\Omega_{Gi}$ is the
cyclotron frequency of the gas ions. The target plasma is
composed of electrons and hexavalent
ions with ion-to-electron mass ratio $m_T/m_e=200$. The relative
density of the target electrons to the gas electrons is
$N_{Te}/N_{Ge}=9$ and the magnetic field
carried by the target plasma is 6 times
the ambient field. The injection speed of the target ions
is $v_{in}=27v_A$, where $v_A$
denotes the Alfv\'{e}n velocity in the gas plasma. The
distribution function of target electrons is given by a
full-Maxwellian in $v_y$ and $v_z$, and a half-Maxwellian
in $v_x (>0)$. The temperature of the target electrons
(and ions) is assumed to be $T_e = 2 m_e v^2_{in}$.

\textcolor{black}{
The Biermann battery effect through laser-target
interactions are thought to generate $\sim 100$ T
magnetic field at the target surface
\cite{yamazaki20,kugland13,gao15}. This field may decay
away from the target. But it is difficult to infer
precisely the magnetic field carried by the expanding
target plasma. \cite{yamazaki20} performed
two-dimensional radiation hydrodynamics simulation to
estimate electron pressure and density of Al plasma
ejected through laser-target interaction under the same
condition (laser intensity and spot radius) as the
experiment here. Their result shows that the Al plasma
carries $\sim$ 100 (10) T magnetic field after 4 (8) ns
near the target surface.
Here, we assume that the magnetic field carried by the
target plasma at the surface is 22.8 T.
The dimensional effect of a toroidal field cannot be
incorporated in the one-dimensional simulation so that we
assume the magnetic field is perpendicular
($z$-direction) to the simulation axis.} The duration of
the injection of the target plasma is equivalent to $1.3$
ns which is the pulse duration of the HIPER laser. The
injection speed roughly reads $\sim 900$ km/s, which is
close to the value obtained by \cite{yamazaki20}.
Although we use unrealistic mass ratio, the mass ratio of
the target ions to the gas ions (=2) is approximately
equal to that of the Al to N ions. We checked that the
following results are little dependent on the
ion-to-electron mass ratio and the frequency ratio,
$\omega_{Ge}/\Omega_{Ge}$. The system size is
$L=25v_A/\Omega_{Gi}$ ($\approx 32$ mm). The number of
grids is 72,000 and the number of superparticles per cell
is 256 for each species.

\begin{figure}[ht]
\includegraphics[clip, width=1.0\columnwidth]{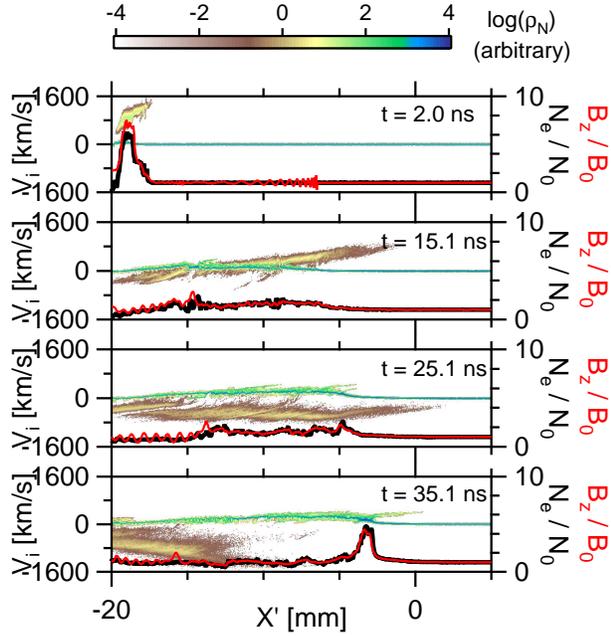}
\caption{\label{fig:PIC_ion_init}\textcolor{black}{Early time evolution of
ion phase space and the profiles of electron density (thick black line)
and magnetic field (red line). The color
scale denotes ion charge density, $\rho_N$.}}
\end{figure}

\begin{figure}[ht]
\includegraphics[clip, width=1.0\columnwidth]{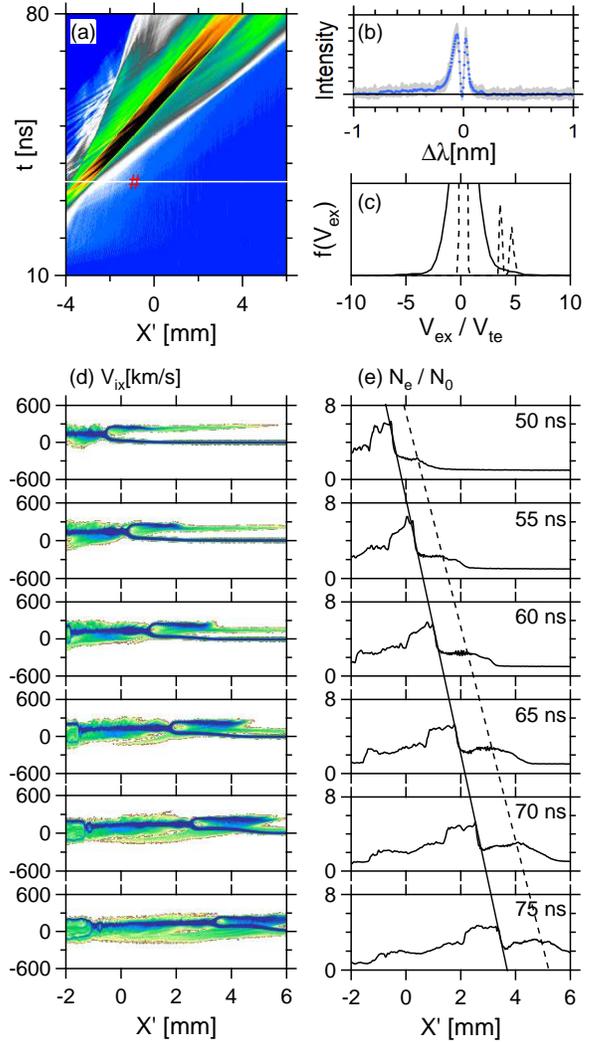}
\caption{\label{fig:PIC}(a) Spatio-temporal evolution of
$N_e$ in the PIC simulation \textcolor{black}{with the
strength of the Biermann field carried by the target plasma
of 22.8 T}. (b) Experimental data of CTS
ion feature for $B=3.8$ T in the precursor at $p=1.36$ mm
and $t=35$ ns. (c) Simulated electron (solid line) and ion
(dashed line) distribution functions at the red sharp in (a).
(d) Time evolution of ion phase space and (e) electron
density profile. The rough positions of the main peak and
the plateau are traced by the solid and the dashed lines,
respectively.}
\end{figure}

Fig.\ref{fig:PIC_ion_init} shows early time evolution of
the ion phase space distribution. Also plotted thick black
line and red line denote the profiles of electron density
and magnetic field, respectively. At $t=2.0$ ns, injected
target ions are easily identified. While they gyrate around
the ambient magnetic field in the gas plasma, they are
elongated in the phase space due to their velocity dispersion
($t = 15.1$ ns). After a quater of their gyro period
($t \approx 19$ ns),
the target ions turned back and
disappeared from the region around the TCC as seen in the
panels of $t=25.1$ ns and 35.1 ns. Because of the short
injection time and the velocity dispersion, the background
gas plasma has not been compressed enough to form a shock
before $t \sim 30$ ns. However, the gas ions are dragged
(or accelerated) by the target ions during a quater of their
gyro motion so that they have positive velocity in $v_x$.
These gas ions are gradually accumulated and compressed
to form a shock-like steepened density profile ($t=35$
ns). However, the density drops immediately behind the
steepened region because no more sweep of the gas plasma
occurs. We call this steepened structure as a \textcolor{black}{developing} shock.

\textcolor{black}{The spatio-temporal evolution of electron density of the
same field of view as the SOP data in
Fig.\ref{fig:sop}(b) is shown in Fig.\ref{fig:PIC}(a). It
shares some common features with the SOP data.} When
$t<40$ ns, a region where the electron density is
slightly enhanced is seen upstream of the main peak.
After $t \sim 40$ ns, \textcolor{black}{in front of the main
peak, the region where the density is a little enhanced grows
(green-colored).} A \textcolor{black}{developing} shock lies at the
sharp boundary just behind the green-colored region. Its
propagation speed at $t = 70$ ns is
$v_{sh} \approx 153$ km/s ($M_A \approx 4.4$), while the propagation
speed of the main peak in the experiment at the same time is
$v_{sh} \approx 146$ km/s ($M_A \approx 3.8$)
[Fig.\ref{fig:sop}(b)].

\textcolor{black}{As mentioned already, the developing shock
shares some characteristics with a fully developed
supercritical shock.}
In Fig.\ref{fig:PIC}(d) the \textcolor{black}{developing} shock stands at $X' \approx
-0.5$ mm when $t=50$ ns. After that, a main peak and a
plateau are clearly seen in the electron density profiles
in Fig.\ref{fig:PIC}(e). Although well developed
uniform downstream has not been realized yet, \textcolor{black}{the spatial extension of the post-ramp exceeds local ion inertial length as well as the thermal gyro radius of main local ion component. Non-negligible amount of incident ions
are reflected at the front of the developing shock, while the}
plateau corresponds to the region where the reflected
ions occupy, i.e., the foot, which extends as the
reflected ions reach farther upstream. Accordingly, the
density profile varies in time. This extension of the
foot is similar to that of the plateau-like structure
seen in the experiment [Fig.\ref{fig:sop}(d)]. The
simulated \textcolor{black}{developing} shock is self-reforming, although the
reformation cycle has not completed by $t=80$ ns.

In the CTS spectrum in the precursor for $B=3.8$ T in the
experiment [Fig.\ref{fig:PIC}(b)] the tail of the
spectrum ($\Delta \lambda < -0.2$ nm) is clearly
enhanced. This enhancement of the blue shifted signals
implies the presence of electrons moving faster than the
bulk toward the upstream. At the similar position and
time in the simulation [red sharp in
Fig.\ref{fig:PIC}(a)], the tail of the electron
distribution function is enhanced for $V_x>0$
[Fig.\ref{fig:PIC}(c)]. Also plotted ion distribution
function (dashed line) shows that some ions having
positive velocity exist, indicating that some electrons
are dragged by these ions.
Therefore, it is inferred that the asymmetric CTS
spectrum is related with the asymmetric electron
distribution function. In the simulation the long time
evolution of the electron distribution function in the
foot after this is clearly correlated with the gyro
motions of reflected ions. Hence, it is expected that the
similar CTS spectrum in an experiment is obtained, if a
well developed foot of a self-reforming shock is formed.

In the end of this section we make some comments. We have
used some unrealistic values of parameters in the
simulation. In particular, ion to electron mass ratios
$m_G /m_e = 100$ and $m_T / m_e = 200$ are quite
smaller than real ($m_N /m_e = 25704$ and $m_{Al} / m_e =
49572$). However, this seems not to lead significant
influence on the results. We have scaled the simulation
parameters in unit of ion quantities, i.e., the
quantities normalized to $c/\omega_{pN}$, $\Omega_{N}$,
and $v_A$ are chosen to match with the experimental data.
This is of course justified only when we focus on the
phenomena controlled by ion dynamics. On the other hand,
there are a number of effects which may cause the
discrepancy from the experiment. For instance, the
spatial expansion of the \textcolor{black}{developing} shock may be a possible reason
for deceleration of its speed
[Fig.\ref{fig:sop}(b)]. Non-uniformity of the system may
also affect the properties of the observed \textcolor{black}{developing} shock in the
experiment. For instance, the ionization level of the gas
plasma may be different between the \textcolor{black}{developing} shock front and far
upstream. All these effects should be taken into account
for more accurate comparison between the simulation and
the experiment.

\begin{figure}[ht]
\includegraphics[clip, width=1.0\columnwidth]{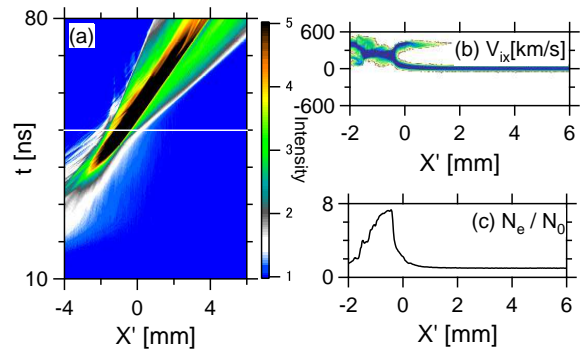}
\caption{\label{fig:PICweakB}(a) Spatio-temporal evolution of
$N_e$ in the PIC simulation with weaker magnetic field of
0.76 T carried
by the target plasma. (b) Ion phase space and (c) electron
density profile at $t=50$ ns. }
\end{figure}

\textcolor{black}{As already mentioned, the estimate of
Biermann battery effect is controversial. Although we
used the value 22.8 T according to \cite{yamazaki20},
there is also an estimate giving much smaller value
of the self-generated magnetic field when the focal
spot is large \cite{ryutov13}.
We have confirmed
that the above result does not change very much
even when the magnetic field carried by the target plasma
is 1/30 times ($\approx 0.76$ T) that of the case we
discussed. Fig.\ref{fig:PICweakB}(a) shows the
spatio-temporal evolution
of $N_e$ in the same format as in Fig.\ref{fig:PIC}(a).
Although the details of instantaneous ion phase space
[Fig.\ref{fig:PICweakB}(b)] and electron density profile
[Fig.\ref{fig:PICweakB}(c)] at $t=50$ ns, for example, are
different from Figs.\ref{fig:PIC}(d) and (e), the feature of
propagation of the developing shock is similar.
This is probably due to that the injected target plasma
diffuses in space due to its velocity dispersion so that
the impact of the magnetic field carried by the target
plasma is not so significant. (Note that we slightly changed
other parameters. The valence of target ions is
5, the relative density of the target electrons to the
gas electrons is $N_{Te}/N_{Ge}=5$, and injection speed
of the target ions is $v_{in}=24 v_A$.)}

\textcolor{black}{\section{summary}}

\textcolor{black}{
In summary, a \textcolor{black}{developing} supercritical collisionless
shock was produced in a uniformly magnetized gas plasma
by using Gekko XII HIPER laser. A long time evolution
of the system after $t > 30$ ns were observed by the SOP
and the CTS measurements. We successfully observed a
\textcolor{black}{developing} shock formed in a homogeneously
magnetized gas plasma without contaminated by the plasma produced
in the vicinity of the solid target in the early time after
the main laser shot. Although the observed \textcolor{black}{developing}
shock has not developed to have uniform downstream, it possesses
some characteristics of a supercritical shock. The
Alfv\'{e}n Mach number exceeds the critical Mach number
($\sim 3$) during the time observed. The width of the
observed transition region varies in time and it is
similar to the feature of the developing foot reproduced
in the PIC simulation.
}

Finally, we emphasize utilities of the high-power laser
experiment. We used an advantage that the information of
space and time is separable to show that the spatial
profile of a \textcolor{black}{developing} shock is time varying. There is also another
advantage, although we have not focused it in this paper.
While a remote sensing used in astrophysics observations
enables us to capture global or macro scale structures of
a phenomenon, local or micro scale structures of the
phenomenon is usually unresolved. In contrast, in-situ
observations in heliospheric physics can access detailed
local information of particles and waves, while the
global structure of the system at the same time is not
accessible. That is, it is difficult in space to
simultaneously measure global/macro and local/micro
structures of a phenomenon. In the experiment one can
simultaneously access multi-scale information. These
indicate that the laser experiment can complement the
conventional observations in space and can be a new basic
tool of the research in high energy astrophysics, space
plasma physics, and other related fields.

\begin{acknowledgments}
The authors would like to acknowledge the dedicated technical support
of the staff at the Gekko-XII facility for the laser operation, target
fabrication, and plasma diagnostics. We would also like to thank M. Hoshino,
T. Hada, Y. Ohira, T. Umeda for useful discussions. This research was
partially supported by the Sumitomo Foundation for environmental research
projects (203099) (SM), JSPS KAKENHI Grant Nos.18H01232, 22H01251 (RY),
17H18270 (SJT), 15H02154 (YS), JSPS Core-to-Core Program B. Asia-Africa
Science Plat- forms Grant No. JPJSCCB20190003 (YS) and by the joint
research project of Institute of Laser Engineering, Osaka University.
The computation was carried out using the computer resource offered
under the category of General Projects by Research Institute for
Information Technology, Kyushu University.
\end{acknowledgments}

\end{document}